\documentclass[prd,twocolumn,showpacs,superscriptaddress,nofootinbib,floatfix,showkeys,10pt]{revtex4-2}

\usepackage{graphicx}
\usepackage{amsmath}
\usepackage{bm}
\usepackage{yhmath}
\usepackage{mathtools}
\usepackage{wasysym}
\usepackage[colorlinks,citecolor=blue,urlcolor=blue,linkcolor=blue]{hyperref}
\usepackage{subfigure}
\usepackage{color}
\usepackage{cases}
\usepackage{subfigure}
\usepackage{times}
\usepackage{dcolumn,booktabs,bm}
\usepackage{slashed}
\usepackage{amsfonts,amssymb,stmaryrd,latexsym,amsmath}
\usepackage{textcomp}
\usepackage{multirow}
\usepackage{cancel}
\usepackage{array}
\usepackage{orcidlink}
\usepackage{enumitem}

\newcommand{\clabel}[2][]{#2}

\newcommand{\etal}{\textit{et al}. }
\renewcommand{\arraystretch}{1.8}

\begin{document}


\title{Doubly heavy tetraquark states in the constituent quark model using diffusion Monte Carlo method}

\author{Yao Ma\,\orcidlink{0000-0002-5868-1166}}\email{yaoma@pku.edu.cn}
\affiliation{School of Physics and Center of High Energy Physics, Peking University, Beijing 100871, China}

\author{Lu Meng\,\orcidlink{0000-0001-9791-7138}}\email{lu.meng@rub.de}
\affiliation{Institut f\"ur Theoretische Physik II, Ruhr-Universit\"at Bochum,  D-44780 Bochum, Germany }

\author{Yan-Ke Chen\,\orcidlink{
0000-0002-9984-163X}}\email{chenyanke@stu.pku.edu.cn}
\affiliation{School of Physics, Peking University, Beijing 100871, China}

\author{Shi-Lin Zhu\,\orcidlink{0000-0002-4055-6906}}\email{zhusl@pku.edu.cn}
\affiliation{School of Physics and Center of High Energy Physics, Peking University, Beijing 100871, China}

\begin{abstract}

We use the diffusion Monte Carlo method to calculate the doubly heavy tetraquark $T_{cc}$ system in two kinds of constituent quark models, the pure constituent quark model AL1/AP1 and 
the chiral constituent quark model. When the discrete configurations are complete and no spatial clustering is preseted, the AL1/AP1 model gives an energy of $T_{cc}$ close to the $DD^*$ threshold, and the chiral constituent quark model yields a deeply bound state. We further calculate all doubly heavy tetraquark systems with $J^P=0^+,1^+,2^+$, and provide the binding energies of systems with bound states. The $I(J^P)=0(0^+)$ $bc\bar{n}\bar{n}$, 
$0(1^+)$ $bb\bar{n}\bar{n}$, 
$0(1^+)$ $bc\bar{n}\bar{n}$,  $\frac{1}{2}(1^+)$ $bb\bar{s}\bar{n}$ systems have bound states in all three models. Since the DMC method has almost no restriction on the spatial part, the resulting bound states have greater binding energies than those obtained in previous works.

\end{abstract}

\maketitle

\section{Introduction}~\label{sec:intro}

The investigations of doubly heavy tetraquark states $QQ\bar{q}\bar{q}$ ($Q=c,b$, $q=u,d,s$) can be traced back to forty years ago. In the 1980s, several works used the non-relativistic quark models to study whether a bound state of $QQ\bar{q}\bar{q}$ exists below the dimeson $Q\bar{q}$ threshold~\cite{Ader:1981db,Zouzou:1986qh,Carlson:1987hh}. It was found that a bound state with $M(QQ\bar{q}\bar{q})<2M(Q\bar{q})$ can always occur when $\frac{m_{Q}}{m_{q}}$ is sufficiently large. The bound state system of $QQ\bar{q}\bar{q}$ has a very characteristic
structure of the baryon type $\bar{Q}'\bar{q}\bar{q}$, reflecting the heavy antiquark-diquark symmetry~\cite{Zouzou:1986qh}. After that, numerous theoretical studies had concentrated on double-heavy tetraquark systems using various approaches, including the color–magnetic interaction model~\cite{Cui:2006mp,Lee:2009rt,Luo:2017eub}, non-relativistic quark model~\cite{Ader:1981db,Zouzou:1986qh,Carlson:1987hh,Silvestre-Brac:1993zem,Semay:1994ht,Pepin:1996id,Vijande:2003ki,Vijande:2007rf,Yang:2009zzp,Park:2018wjk,Deng:2018kly,Maiani:2019lpu,Yang:2019itm,Tan:2020ldi,Meng:2020knc,Noh:2021lqs}, relativized quark model~\cite{Lu:2020rog}, relativistic quark model~\cite{Ebert:2007rn,Faustov:2021hjs}, hadronic molecular picture~\cite{Janc:2004qn,Li:2012ss,Xu:2017tsr,Liu:2019stu}, QCD sum rule~\cite{Navarra:2007yw,Dias:2011mi,Du:2012wp,Wang:2017uld,Gao:2020ogo}, lattice QCD~\cite{Ikeda:2013vwa,Cheung:2017tnt,Francis:2018jyb,Junnarkar:2018twb,Hudspith:2020tdf}, heavy-quark symmetry~\cite{Eichten:2017ffp,Braaten:2020nwp,Cheng:2020wxa} and others~\cite{Gelman:2002wf,Karliner:2017qjm}.

In 2021, the LHCb collaboration observed the first doubly charmed tetraquark state $T_{cc}^+(3875)$ in the $D^0D^0\pi^+$ mass spectrum~\cite{LHCb:2021vvq,LHCb:2021auc}. Its mass is very close to the $D^{*+}D^0$ threshold, with a binding energy of only about 300 keV. It is a state with quark composition $cc\bar{n}\bar{n}$ ($n=u,d$) favoring the quantum number $J^P=1^+$. The experimental findings have sparked renewed interest in doubly heavy tetraquark states~\cite{Meng:2021jnw,Feijoo:2021ppq,Dong:2021bvy,Chen:2021tnn,Chen:2021cfl,Du:2021zzh,Deng:2021gnb,Chen:2021spf,Deng:2022cld,Cheng:2022qcm,Wang:2022jop,He:2022rta,Wang:2023ovj,Du:2023hlu,Lyu:2023xro,Wang:2023iaz,Dai:2023kwv,Meinel:2022lzo,Hudspith:2023loy,Padmanath:2023rdu}. One can see more comprehensive reviews in Refs.~\cite{Lebed:2016hpi,Chen:2016qju,Guo:2017jvc,Liu:2019zoy,Chen:2022asf,Meng:2022ozq}. The discovery of $T_{cc}^+$ sheds light on the research of multiquark states, and suggests the possible existence of other doubly heavy tetraquark states.

A minimal constituent quark model used to describe multi-quark states typically incorporates the one-gluon-exchange interaction and the confinement effect. Some quark models may also encompass additional interactions, such as one-boson exchange (OBE) interactions. They include \clabel[pseudoscalar]{pseudoscalar} meson exchange interactions stemming from the spontaneous breaking of chiral symmetry, vector meson exchange interactions, and scalar meson exchange interactions. Based on the inclusion or exclusion of the OBE interaction, constituent quark models can be broadly categorized into two types: the pure constituent quark model (PCQM) without OBE and the chiral constituent quark model ($\chi$CQM) with OBE. One of the objectives of this study is to investigate the difference between these two types of quark model interactions concerning their predictions of the doubly heavy tetraquark states.

\begin{figure}[htbp]
  \centering
  \includegraphics[width=0.47\textwidth]{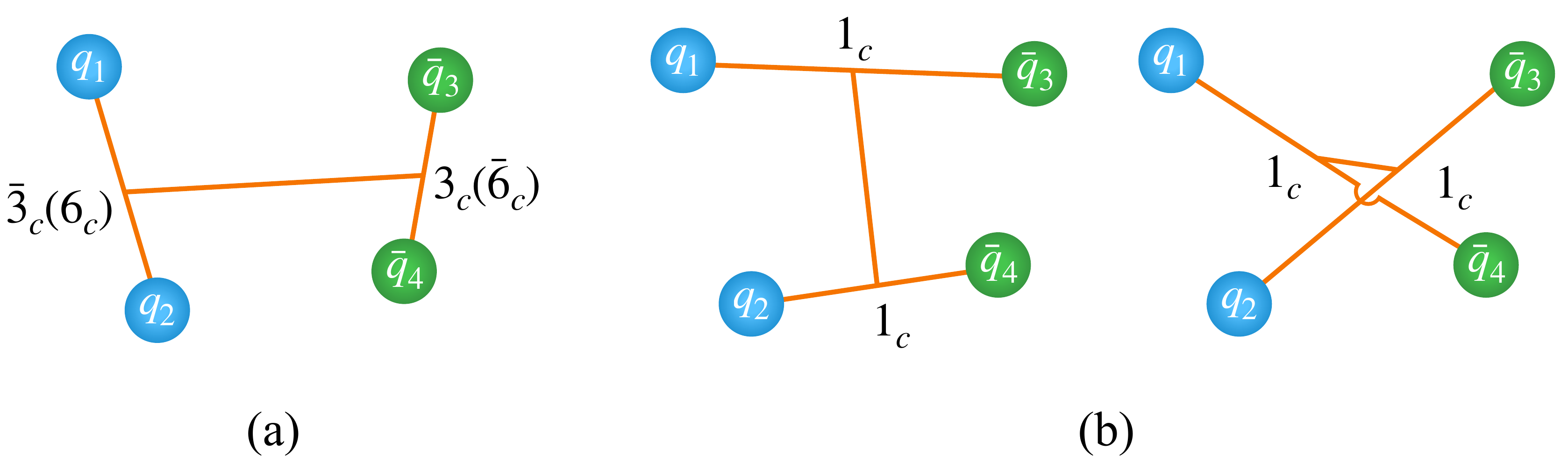} 
  \caption{\label{fig:structure} Two structures for the tetraquark system. (a) diquark-antidiquark structure. (b) dimeson structure. }
    \setlength{\belowdisplayskip}{1pt}
\end{figure}

With the quark interactions, one can solve the four-body \clabel[SchEq]{Schr\"odinger equation} based on the variational method via the basis expansions. Apparently, the precision of the solution depends on the choice of trial functions or basis functions.  In the practical calculation, the diquark-antidiquark structure and the dimeson structure are two widely used structures to investigate tetraquark states. The diquark-antidiquark structure has a color wave function where two quarks (antiquarks) combine first and then form a color singlet together, namely $[(qq)_{\overline{3}_{c}}(\bar{q}\bar{q})_{3_{c}}]_{1_{c}}$ and $[(qq)_{6_{c}}(\bar{q}\bar{q})_{\bar{6}_{c}}]_{1_{c}}$. In the spatial part, one can choose the corresponding Jacobi coordinates to depict the correlation as shown in Fig.~\ref{fig:structure}, where the diquark and antidiquark form two clusters separately and then form the tetraquark state. The dimeson structure usually restricts the color wave function to be $[(q\bar{q})_{1_{c}}(q\bar{q})_{1_{c}}]_{1_{c}}$. It should be noticed that even if the $[(q\bar{q})_{8_{c}}(q\bar{q})_{8_{c}}]_{1_{c}}$ is not included, the color bases with two dimeson structures by exchanging (anti)quarks is complete, though they are not orthogonal. In principle, one could choose different structures for wave functions of discrete quantum numbers and spatial wave functions. For example, the spatial wave functions are the dimeson structure, while allowing the color part to contain both $[(q\bar{q})_{1_{c}}(q\bar{q})_{1_{c}}]_{1_{c}}$ and $[(q\bar{q})_{8_{c}}(q\bar{q})_{8_{c}}]_{1_{c}}$. If it is such a case, we will specify it explicitly.

In previous investigations of the $cc\bar{n}\bar{n}$ system with $(I,J)=(0,1)$, it has been noted that the choice of different structural configurations yields disparate outcomes.  For a PCQM, Yang \etal obtained a shallow bound state under the dimeson structure, while they found no bound state under the diquark-antidiquark structure~\cite{Yang:2009zzp}. For the $\chi$CQM, they obtained a shallow bound state under the dimeson structure and a deeply bound state under the diquark-antidiquark structure. Similar discrepancy was also observed in other $\chi$CQM works~\cite{Yang:2019itm,Chen:2021tnn,Deng:2021gnb,Deng:2022cld}. In other words, to accurately determine the ground state energies of the multiquark states using a variational method-based approach, one must either conceive a good conjecture about the clustering behavior of the wave functions or employ very general trial functions. Alternatively, the diffusion Monte Carlo (DMC) method is a promising technique to precisely and efficiently ascertain ground state energies, without the need for a priori assignment of clustering behaviors.

DMC has been well used in the simulations of the molecular physics~\cite{suhm1991quantum}, solid physics~\cite{Foulkes:2001zz}, and nuclear physics~\cite{Carlson:2014vla}. In hadronic physics, the DMC method has been used in quark models in several works~\cite{Bai:2016int,Gordillo:2020sgc,Gordillo:2021bra,Alcaraz-Pelegrina:2022fsi,Gordillo:2022nnj,Ma:2022vqf,Gordillo:2023tnz}. In the DMC formalism, the distribution of the so-called walkers is used to represent the spatial wave function, which gradually evolves to the ground state over time in principle. The DMC approach does not select a specific basis for the spatial wave function, resulting in a general wave function space. This approach could be more suitable for exploring the discrepancy caused by predetermined clustering structures.

This paper is arranged as follows. 
In Sec.~\ref{sec:DMC}, the diffusion Monte Carlo method is introduced. 
In Sec.~\ref{sec:Hamiltonian}, the Hamiltonians with different types of potential models are presented. In Sec.~\ref{sec:wavefunction}, the construction of the wave function for all doubly heavy tetraquark systems with $J^P=0^+,1^+,2^+$ is presented.
In Sec.~\ref{sec:results}, we give the numerical results for the $T_{cc}$ system in different models, and the bound states of all other doubly heavy tetraquark systems with $J^P=0^+,1^+,2^+$. 
Finally, a brief discussion and summary are given in Sec.~\ref{sec:sum}.

\section{Diffusion Monte Carlo method}~\label{sec:DMC}

We adhere to the formalism presented in Ref.~\cite{Ma:2022vqf}. The DMC algorithm can be derived from the imaginary-time Schr\"odinger equation (in natural units $\hbar=c=1$) and its subsequent solution.
\begin{align}\label{schrodinger}
-\frac{\partial\Psi(\boldsymbol{R},t)}{\partial t}&=[\hat{H}-E_{R}]\Psi(\boldsymbol{R},t)\,,\\
\hat{H}&=-\sum_{i=1}^m \frac{1}{2m_i}\boldsymbol{\nabla}_{\boldsymbol{r}_{i}}^{2}+V(\boldsymbol{R})\,,\\
\Psi(\boldsymbol{R},t)&=\sum_{i}c_{i}\Phi_{i}(\boldsymbol{R})e^{-[E_{i}-E_{R}]t}\,,
\end{align}
where  $\boldsymbol{R}\equiv(\boldsymbol{r}_1,\boldsymbol{r}_2,...,\boldsymbol{r}_m)$ represents the positions of $m$ particles. $E_{R}$ is the shift of energy, and $\Phi_{i}(\boldsymbol{R})$ is the eigenfunctions. When $E_{R}$ is taken close to the ground state energy $E_0$, the wave function $\Psi(\boldsymbol{R},t)$ will approach $\Phi_{0}(\boldsymbol{R})$ after a sufficiently long time evolution (as long as $c_0$ is not too small), and other components will be suppressed exponentially~\cite{Boronat1994}.

The Eq.~\eqref{schrodinger} has a solution in the form of path integral,
\begin{align}\label{path_int}
\Psi(\boldsymbol{R}', t+\Delta t) 
\approx\int &d \boldsymbol{R}_1 d\boldsymbol{R}_2 d \boldsymbol{R} \ G_0\left(\boldsymbol{R}', \boldsymbol{R}_1, \frac{\Delta t}{2}\right) \nonumber\\
&\times G_1\left(\boldsymbol{R}_1, \boldsymbol{R}_2, \Delta t\right) G_0\left(\boldsymbol{R}_2, \boldsymbol{R}, \frac{\Delta t}{2}\right) \nonumber\\
&\times\Psi\left(\boldsymbol{R}, t\right) \,,
\end{align}
where

\begin{align}\label{Green}
G_{1}(\boldsymbol{R}',\boldsymbol{R},t)&=\prod_{i=1}^m\left(\frac{2\pi t}{m_{i}}\right)^{-3/2} e^{-\frac{m_{i}}{2t}(\boldsymbol{r}'_{i}-\boldsymbol{r}_{i})^{2}}\,,\nonumber\\
G_{0}(\boldsymbol{R}', \boldsymbol{R}, t)&=e^{-\left(V(\boldsymbol{R})-E_R\right) t
}\delta \left(\boldsymbol{R}'-\boldsymbol{R}\right)\,,
\end{align}
Theoretically, this form of the solution can be readily implemented using an algorithm. In this algorithm, a substantial number of walkers sample the wave function $\Psi(\boldsymbol{R},t)$ and the distribution of walkers represents the amplitude of $\Psi(\boldsymbol{R},t)$. 

However, this algorithm is usually unstable due to the drastic statistic fluctuation. We need to further use the importance sampling technique to reduce the fluctuation~\cite{Kalos1974}. In this approach, rather than directly sampling the wave function $\Psi(\boldsymbol{R},t)$, we sample a newly defined function, denoted as $f(\boldsymbol{R},t)$,
\begin{equation}\label{f}
f(\boldsymbol{R},t)\equiv\psi_T(\boldsymbol{R})\Psi(\boldsymbol{R},t)\,,
\end{equation}
where $\psi_T(\boldsymbol{R})$ is a time-independent importance function, which should be chosen as close to $\Phi_{0}(\boldsymbol{R})$ as possible. 
In this work, we use the form following Ref.~\cite{Gordillo:2020sgc},
\begin{equation}\label{psi_T}
\psi_T(\boldsymbol{R})=\prod_{i<j} e^{-a_{ij}r_{ij}}\,.
\end{equation}
Here $a_{ij}$ are adjustable constants and their values are set to minimize the fluctuation. 

The Schr\"odinger equation with importance sampling reads
\begin{align}\label{schrodingerForf}
-\frac{\partial f(\boldsymbol{R},t)}{\partial t}=&-\sum_{i=1}^m \frac{1}{2m_i}\boldsymbol{\nabla}_{\boldsymbol{r}_{i}}^{2}f(\boldsymbol{R},t)\nonumber\\
&+\sum_{i=1}^m\frac{1}{2m_i}\boldsymbol{\nabla}_{\boldsymbol{r}_{i}}(\boldsymbol{F}_i(\boldsymbol{R})f(\boldsymbol{R},t))\nonumber\\
&+[E_{L}(\boldsymbol{R})-E_{R}]f(\boldsymbol{R},t)\,,
\end{align}
where $E_{L}(\boldsymbol{R})=\psi_T(\boldsymbol{R})^{-1}\hat{H}\psi_T(\boldsymbol{R})$ is the local energy, $\boldsymbol{F}_i(\boldsymbol{R})=2\psi_T(\boldsymbol{R})^{-1}\boldsymbol{\nabla}_{\boldsymbol{r}_i}\psi_T(\boldsymbol{R})$ is the drift force acting on particle $i$. The path integral solution of Eq.~\eqref{schrodingerForf} is
\begin{align}\label{evolution}
f(\boldsymbol{R}',t+\Delta t)
\approx&\int \mathrm{d}\boldsymbol{R_1}\mathrm{d}\boldsymbol{R_2}\mathrm{d}\boldsymbol{R_3}\mathrm{d}\boldsymbol{R_4}\mathrm{d}\boldsymbol{R}\nonumber\\
&\times G_{3}(\boldsymbol{R}',\boldsymbol{R_{1}},\frac{\Delta t}{2})G_{2}(\boldsymbol{R_{1}},\boldsymbol{R_{2}},\frac{\Delta t}{2})\nonumber\\
&\times G_{1}(\boldsymbol{R_{2}},\boldsymbol{R_{3}},\Delta t)\nonumber\\
&\times G_{2}(\boldsymbol{R_{3}},\boldsymbol{R_{4}},\frac{\Delta t}{2})G_{3}(\boldsymbol{R_{4}},\boldsymbol{R},\frac{\Delta t}{2})\nonumber\\
&\times f(\boldsymbol{R},t)\,,
\end{align}
with
\begin{eqnarray}
    G_{1}(\boldsymbol{R}',\boldsymbol{R},t)&=&\prod_{i=1}^m\left(\frac{2\pi t}{m_{i}}\right)^{-3/2}e^{-\frac{m_{i}}{2t}(\boldsymbol{r}'_{i}-\boldsymbol{r}_{i})^{2}}\,, \nonumber\\
    G_{2}(\boldsymbol{R}',\boldsymbol{R},t)&=&\prod_{i=1}^m\delta\left(\boldsymbol{r}'_{i}-\boldsymbol{r}_{i}-\frac{\boldsymbol{F}_{i}(\boldsymbol{R})}{2m_{i}}t\right)\,, \nonumber\\
    G_{3}(\boldsymbol{R}',\boldsymbol{R},t)&=&e^{-(E_{L}(\boldsymbol{R})-E_{R})t}\delta \left(\boldsymbol{R}'-\boldsymbol{R}\right)\,.
\end{eqnarray}
The DMC algorithm employed to obtain the solution described above, along with the formalism utilized to address the coupled-channel, has been comprehensively elucidated in Ref.~\cite{Ma:2022vqf}.

\section{Hamiltonian}~\label{sec:Hamiltonian}

The nonrelativistic Hamiltonian of a four-quark system reads
\begin{align}
H=\sum_i^4\left(m_i+\frac{\boldsymbol{p}_i^2}{2m_i}\right)-T_{CM}+V\,,
\end{align}
where $m_i$ and $\boldsymbol{p}_i$ are the mass and momentum of quark $i$. $T_{CM}$ is the center-of-mass kinematic energy, which automatically vanishes in the evolution, because the system will tend to the lowest energy state.

In this study, we will employ two types of non-relativistic quark models: pure constituent quark model (PCQM) and chiral constituent quark model ($\chi$CQM). The primary distinction between these two types is that the $\chi$CQM incorporates the OBE between quarks, a feature not included in the PCQM.

For the PCQM, we adopt AL1 and AP1 quark models~\cite{Semay:1994ht,Silvestre-Brac:1996myf}, which contain the one-gluon-exchange (OGE) interaction and the confinement effect.
The potential reads 
\begin{align}\label{AL1}
    V_{i j}^{pure}=-\frac{3}{16}&\lambda_i^c \cdot \lambda_j^c\Big(-\frac{\kappa}{r_{ij}}+\lambda r_{ij}^p-\Lambda\nonumber\\
    &+\frac{8\pi\kappa'}{3m_{i}m_{j}}\frac{\exp(-r_{ij}^{2}/r_{0}^{2})}{\pi^{3/2}r_{0}^{3}}\boldsymbol{s}_{i}\cdot\boldsymbol{s}_{j}
    \Big),
\end{align}
where $r_0=A\left(\frac{2m_im_j}{m_i+m_j}\right)^{-B}$. The power parameter $p$ of the confinements are $1$ and $2/3$ for AL1 and AP1, respectively. $\lambda^c$ represents the SU(3) color Gell-Mann matrix. $\boldsymbol{s}_i$ is the spin operator of quark $i$. The parameters of the two models are taken from Ref.~\cite{Silvestre-Brac:1996myf}. There are some similar models, such as Barnes, Godefrey and Swanson model~\cite{Barnes:2005pb}. However, the parameters for the light quarks were not determined.

For the $\chi$CQM, we refer to the model introduced in Ref.~\cite{Vijande:2004he,Segovia:2011dg} as the Salamanca model (SLM), which encompasses the OBE interaction, OGE interaction, and a confinement potential with a screening effect,
\begin{align}
V_{i j}^{chiral}&=V^{\mathrm{OBE}}_{i j}+V^{\mathrm{OGE}}_{i j}+V^{\mathrm{CON}}_{i j}\,.
\end{align}
In this work, only the central term is considered. 

The OGE potential includes a Coulomb term and a spin-dependent color-magnetic term,
\begin{align}
V^{\mathrm{OGE}}_{i j}=&\frac{1}{4} \alpha_s^{ij} \lambda_i^c \cdot \lambda_j^c\nonumber\\
&\times\left[\frac{1}{r_{i j}}-\frac{1}{6 m_i m_j} \boldsymbol{\sigma}_i \cdot \boldsymbol{\sigma}_j \frac{\mathrm{e}^{-r_{i j} / r_0(\mu_{ij})}}{r_{i j} r_0^2(\mu_{ij})}\right]\,,
\end{align}
where $r_0(\mu_{ij})= \frac{\hat{r}_0}{\mu_{i j}}$, $\mu_{i j}$ is the reduced mass of quark $i$ and $j$. $\lambda^c$ represents the SU(3) color Gell-Mann matrix. The effective scale-dependent strong coupling constant $\alpha_s^{ij}$ is given by
\begin{equation}
\alpha_s^{ij}=\frac{\alpha_0}{\ln \left(\frac{\mu_{ij}^2+\mu_0^2}{\Lambda_0^2}\right)}\,.
\end{equation}

The confinement part is
\begin{equation}
V^{\mathrm{CON}}_{i j}= \lambda_i^c \cdot \lambda_j^c\left[-a_c\left(1-\mathrm{e}^{-\mu_c r_{i j}}\right)+\Delta\right]\,.
\end{equation}
This screened form is introduced to simulate the string breaking effect~\cite{Vijande:2004he}. It exhibits a linear behavior at short distances and becomes constant at large distances.

The meson exchange interactions only occur between the light quarks $q=u,d,s$. The OBE interactions read
\begin{align}
V^{\mathrm{OBE}}_{i j}=& 
\ V^\pi_{i j}\sum_{a=1}^3\left(\lambda_i^a \cdot \lambda_j^a\right)+V^K_{i j} \sum_{a=4}^7\left(\lambda_i^a \cdot \lambda_j^a\right)\nonumber\\
&+V^\eta_{i j}\left[\cos \theta_P\left(\lambda_i^8 \cdot \lambda_j^8\right)-\sin \theta_P\right]\nonumber\\
&+V^\sigma_{i j}\,,
\end{align}
where
\begin{align}
V^\chi_{i j}= & \frac{g_{\mathrm{ch}}^2}{4 \pi} \frac{m_\chi^2}{12 m_i m_j} \frac{\Lambda_\chi^2}{\Lambda_\chi^2-m_\chi^2} m_\chi\left(\boldsymbol{\sigma}_i \cdot \boldsymbol{\sigma}_j\right)\nonumber\\
&\times\left[Y\left(m_\chi r_{i j}\right)-\frac{\Lambda_\chi^3}{m_\chi^3} Y\left(\Lambda_\chi r_{i j}\right)\right]\,,\ \chi=\pi,K,\eta\,,\label{OPE}\\
V^\sigma_{i j}= & -\frac{g_{\mathrm{ch}}^2}{4 \pi} \frac{\Lambda_\sigma^2}{\Lambda_\sigma^2-m_\sigma^2} m_\sigma\left[Y\left(m_\sigma r_{i j}\right)-\frac{\Lambda_\sigma}{m_\sigma} Y\left(\Lambda_\sigma r_{i j}\right)\right]\,.\nonumber
\end{align}
$Y(x)$ is the Yukawa function $Y(x)=\mathrm{e}^{-x} / x$. $r_{ij}$ is the relative distance between quark $i$ and $j$. $\sigma_i$ is the spin Pauli matrix for quark $i$. $\lambda^a$ represents the SU(3) flavor Gell-Mann matrix. For interaction between $q_i\bar{q}_j$, $\lambda_i^a\cdot\lambda_j^a$ should be replaced by $\lambda_i^a \cdot\left(\lambda_j^a\right)^T$. $m_\pi$, $m_K$, and $m_\eta$ are the pseudoscalar meson masses, and $m_\sigma$ is determined through $m_\sigma^2\sim m_\pi^2+4m_{u,d}^2$ as suggested by Ref.~\cite{Scadron:1982eg}. The angle $\theta_P$ is introduced to obtain the physical $\eta$ meson instead of the flavor octet state $\eta_8$. The chiral coupling constant $g_{\mathrm{ch}}$ can be determined from the $\pi N N$ coupling constant through
\begin{equation}
\frac{g_{\mathrm{ch}}^2}{4 \pi}=\left(\frac{3}{5}\right)^2 \frac{g_{\pi \mathrm{NN}}^2}{4 \pi} \frac{m_{u, d}^2}{m_N^2}.
\end{equation}
The parameters of SLM listed in Table ~\ref{ChiralSegovia} are taken from Ref.~\cite{Segovia:2011dg}, which are fitted over all meson spectra.

\begin{table}[htbp] 
\renewcommand{\arraystretch}{1.4}
\centering
\caption{\label{ChiralSegovia} SLM parameters~\cite{Segovia:2011dg}.}
\begin{tabular*}{\hsize}{@{}@{\extracolsep{\fill}}lcc@{}}

\hline\hline
Quark masses & $m_n\text{ }(\mathrm{MeV})$ & 313 \\
& $m_s\text{ }(\mathrm{MeV})$ & 555 \\
& $m_c\text{ }(\mathrm{MeV})$ & 1763 \\
& $m_b\text{ }(\mathrm{MeV})$ & 5110 \\
Goldstone bosons & $m_\pi\text{ }\left(\mathrm{fm}^{-1}\right)$ & 0.70\\
& $m_\sigma\text{ }\left(\mathrm{fm}^{-1}\right)$ & 3.42\\
& $m_K\text{ }\left(\mathrm{fm}^{-1}\right)$ & 2.51 \\
& $m_\eta\text{ }\left(\mathrm{fm}^{-1}\right)$ & 2.77\\
& $\Lambda_\pi\text{ }\left(\mathrm{fm}^{-1}\right)$ & 4.20 \\
& $\Lambda_\sigma\text{ }\left(\mathrm{fm}^{-1}\right)$ & 4.20\\
& $\Lambda_K\text{ }\left(\mathrm{fm}^{-1}\right)$ & 4.21 \\
& $\Lambda_\eta\text{ }\left(\mathrm{fm}^{-1}\right)$ & 5.20 \\
& $g_{c h}^2 / 4 \pi$ & 0.54\\
& $\theta_p\text{ }\left({ }^{\circ}\right)$ & $-15$ \\
OGE & $\alpha_0$ & 2.118  \\
& $\Lambda_0\text{ }\left(\mathrm{fm}^{-1}\right)$ & 0.113\\
& $\mu_0\text{ }\left(\mathrm{MeV}\right)$ & 36.976\\
& $\hat{r}_0\text{ }(\mathrm{MeV} \cdot \mathrm{fm})$ & 28.327\\
Confinement & $a_c\text{ }(\mathrm{MeV})$ & 507.4 \\
& $\mu_c\text{ }\left(\mathrm{fm}^{-1}\right)$ & 0.576\\
& $\Delta\text{ }(\mathrm{MeV})$ & 184.432 \\
\hline \hline
\end{tabular*}
\end{table}

\clabel[meson]{The calculated meson masses in AL1, AP1 and SLM models are shown in Table~\ref{tab:meson}. It can be seen that these potential models can describe the experimental meson spectra well.}

\begin{table}[htbp] 
\renewcommand{\arraystretch}{1.4}
\centering
\caption{\label{tab:meson}The masses (in MeV) of mesons in AL1, AP1 and SLM models, compared with the experimental results taken
from Ref.~\cite{ParticleDataGroup:2022pth}. The experimental results are averaged over the isospin multiples. }
\begin{tabular*}{\hsize}{@{}@{\extracolsep{\fill}}lccccc@{}}

\hline\hline
Mesons& $m_{\mathrm{Exp.}}$& $m_{\mathrm{AL1}}$& $m_{\mathrm{AP1}}$& $m_{\mathrm{SLM}}$&\\
\hline
$B_s$&5367&5362&5356&5348&\\
$B$&5279&5294&5312&5275&\\
$D_s$&1968&1964&1955&1984&\\
$D$&1867&1863&1883&1896&\\
$B_s^*$&5415&5417&5419&5395&\\
$B^*$&5325&5351&5368&5319&\\
$D_s^*$&2112&2102&2108&2112&\\
$D^*$&2009&2017&2034&2019&\\

\hline \hline
\end{tabular*}
\end{table}

\section{Wave function construction}\label{sec:wavefunction}

The experimental $T_{cc}^+(3875)$~\cite{LHCb:2021vvq,LHCb:2021auc} is a candidate of the $cc\bar{n}\bar{n}$ ($n$=$u$ and $d$) tetraquark state with quantum number  $(I)J^P=(0)1^+$. We sequentially label the (anti-)quarks of $cc\bar{n}\bar{n}$  as $1,2,3,4$. The two heavy quarks, designated as $1$ and $2$, are indistinguishable, as are the two light quarks, labeled as $3$ and $4$. The construction of the wave function needs to fulfill the Pauli principle. For $I=0$, the flavor wave function is symmetric for $cc$, while antisymmetric for $\bar{n}\bar{n}$. The remaining color-spin-spatial wave functions allowed by the Pauli principle are
\begin{align}
|T_1^0\rangle &=\left[(12)_{\bar{3}_{c}}^{1_{s}}(34)_{3_{c}}^{0_{s}}\right]_{1_{c}}^{1_{s}}\psi_1^{SS}(12;34),\\
|T_2^0\rangle &=\left[(12)_{6_{c}}^{0_{s}}(34)_{\bar{6}_{c}}^{1_{s}}\right]_{1_{c}}^{1_{s}}\psi_2^{SS}(12;34),\\
|T_3^0\rangle &=\left[(12)_{\bar{3}_{c}}^{1_{s}}(34)_{3_{c}}^{1_{s}}\right]_{1_{c}}^{1_{s}}\psi_3^{SA}(12;34),\\
|T_4^0\rangle &=\left[(12)_{6_{c}}^{1_{s}}(34)_{\bar{6}_{c}}^{1_{s}}\right]_{1_{c}}^{1_{s}}\psi_4^{AS}(12;34),\\
|T_5^0\rangle &=\left[(12)_{\bar{3}_{c}}^{0_{s}}(34)_{3_{c}}^{1_{s}}\right]_{1_{c}}^{1_{s}}\psi_5^{AA}(12;34),\\
|T_6^0\rangle &=\left[(12)_{6_{c}}^{1_{s}}(34)_{\bar{6}_{c}}^{0_{s}}\right]_{1_{c}}^{1_{s}}\psi_6^{AA}(12;34)\,,
\end{align}
where $S$ represents exchange symmetric, and $A$ represents exchange antisymmetric. For the system with isospin $I=1$, the flavor wave function for $\bar{n}\bar{n}$ is symmetric. The remaining color-spin-spatial wave functions need to be changed accordingly
\begin{align}
|T_1^1\rangle &=\left[(12)_{\bar{3}_{c}}^{1_{s}}(34)_{3_{c}}^{1_{s}}\right]_{1_{c}}^{1_{s}}\phi_1^{SS}(12;34),\\
|T_2^1\rangle &=\left[(12)_{6_{c}}^{1_{s}}(34)_{\bar{6}_{c}}^{1_{s}}\right]_{1_{c}}^{1_{s}}\phi_2^{AA}(12;34),\\
|T_3^1\rangle &=\left[(12)_{\bar{3}_{c}}^{1_{s}}(34)_{3_{c}}^{0_{s}}\right]_{1_{c}}^{1_{s}}\phi_3^{SA}(12;34),\\
|T_4^1\rangle &=\left[(12)_{6_{c}}^{1_{s}}(34)_{\bar{6}_{c}}^{0_{s}}\right]_{1_{c}}^{1_{s}}\phi_4^{AS}(12;34),\\
|T_5^1\rangle &=\left[(12)_{\bar{3}_{c}}^{0_{s}}(34)_{3_{c}}^{1_{s}}\right]_{1_{c}}^{1_{s}}\phi_5^{AS}(12;34),\\
|T_6^1\rangle &=\left[(12)_{6_{c}}^{0_{s}}(34)_{\bar{6}_{c}}^{1_{s}}\right]_{1_{c}}^{1_{s}}\phi_6^{SA}(12;34).
\end{align}

\begin{table*}[htp]
    \centering
    \caption{The doubly heavy tetraquark states explored in this study and their associated dimeson thresholds. }
    \label{threshold}
\begin{tabular*}{\hsize}{@{}@{\extracolsep{\fill}}ccccc@{}}
\hline 
\hline 
\multirow{2}{*}{System} & \multirow{2}{*}{$I$} & \multicolumn{3}{c}{Thresholds}\tabularnewline
\cline{3-5} 
 &  & $0^{+}$ & $1^{+}$ & $2^{+}$\tabularnewline
\hline 
\multirow{2}{*}{$cc\bar{n}\bar{n}$} & 0 & $(D_1D)_P/(D_0^*D^*)_P$~\footnote{There is no relevant S-wave thresholds of two ground state mesons in this quantum numbers considering the exchange symmetry of two identical bosons. The subscript represents the relative angular momentum is P-wave.

} & $DD^{*}$ & $(D_1D)_P/(D_0^*D^*)_P$ \tabularnewline
 & 1 & $DD$ & $DD^{*}$ & $D^{*}D^{*}$\tabularnewline
$cc\bar{s}\bar{n}$ & $\frac{1}{2}$ & $DD_{s}$ & $D^{*}D_{s}/D_{s}^{*}D$ & $D^{*}D_{s}^{*}$\tabularnewline
$cc\bar{s}\bar{s}$ & 0 & $D_{s}D_{s}$ & $D_{s}D_{s}^{*}$ & $D_{s}^{*}D_{s}^{*}$\tabularnewline
\hline 
\multirow{2}{*}{$bb\bar{n}\bar{n}$} & 0 &$(\bar{B}_1\bar{B})_P/(\bar{B}_0^*\bar{B}^*)_P$  & $\bar{B}\bar{B}^{*}$ & $(\bar{B}_1\bar{B})_P/(\bar{B}_0^*\bar{B}^*)_P$  \tabularnewline
 & 1 & $\bar{B}\bar{B}$ & $\bar{B}\bar{B}^{*}$ & $\bar{B}^{*}\bar{B}^{*}$\tabularnewline
$bb\bar{s}\bar{n}$ & $\frac{1}{2}$ & $\bar{B}\bar{B}_{s}$ & $\bar{B}_{s}^{*}\bar{B}/\bar{B}^{*}\bar{B}_{s}$ & $\bar{B}^{*}\bar{B}_{s}^{*}$\tabularnewline

$bb\bar{s}\bar{s}$ & 0 & $\bar{B}_{s}\bar{B}_{s}$ & $\bar{B}_{s}\bar{B}_{s}^{*}$ & $\bar{B}_{s}^{*}\bar{B}_{s}^{*}$\tabularnewline
\hline 
\multirow{2}{*}{$bc\bar{n}\bar{n}$} & 0 & $D\bar{B}$ & $D^{*}\bar{B}/D\bar{B}^{*}$ & $D^{*}\bar{B}^{*}$\tabularnewline
 & 1 & $D\bar{B}$ & $D^{*}\bar{B}/D\bar{B}^{*}$ & $D^{*}\bar{B}^{*}$\tabularnewline
$bc\bar{s}\bar{n}$ & $\frac{1}{2}$ & $\bar{B}_{s}D/D_{s}\bar{B}$ & $D_{s}^{*}\bar{B}/D_{s}\bar{B}^{*}/D^{*}\bar{B}_{s}/D\bar{B}_{s}^{*}$ & $D_{s}^{*}\bar{B}^{*}/\bar{B}_{s}^{*}D^{*}$\tabularnewline
$bc\bar{s}\bar{s}$ & 0 & $\bar{B}_{s}D_{s}$ & $D_{s}^{*}\bar{B}_{s}/D_{s}\bar{B}_{s}^{*}$ & $\bar{B}_{s}^{*}D_{s}^{*}$\tabularnewline
\hline 
\hline
\end{tabular*}
\end{table*}

In addition to the $T_{cc}$ states, we will calculate the ground states of all doubly heavy tetraquark systems with $J^P=0^+,1^+,2^+$. The two heavy quarks could be $bb$, $cc$ and $bc$. The two light antiquarks could be $\bar{n}\bar{n}$, $\bar{s}\bar{s}$ and $\bar{s}\bar{n}$. In Table~\ref{threshold}, we list them specifically and their corresponding dimeson thresholds. We use the same methods to construct the wave functions satisfying the Pauli principle.  The spin-color configuration channels included in our calculation are listed in Table~\ref{SpinColorConf}. For each channel, the exchange symmetry of its spatial part  should be determined according to the quark composition (and isospin) of the system. In contrast to other frameworks, such as those discussed in Refs.~\cite{Meng:2019ilv,Meng:2021kmi}, the calculations do not presume the clustering behavior of quarks. 

\begin{table*}[htbp] 
\centering
\caption{\label{SpinColorConf} The spin-color channels included for all $J^P=0^+,1^+,2^+$. The spin part notations are $\chi^{AA}_0=\left[(12)^{0_s}(34)^{0_s} \right]^{0_s}$, $\chi^{SS}_0=\left[(12)^{1_s}(34)^{1_s} \right]^{0_s}$, $\chi^{SS}_1=\left[(12)^{1_s}(34)^{1_s} \right]^{1_s}$, $\chi^{SA}_1=\left[(12)^{1_s}(34)^{0_s} \right]^{1_s}$,
$\chi^{AS}_1=\left[(12)^{0_s}(34)^{1_s} \right]^{1_s}$,
$\chi^{SS}_2=\left[(12)^{1_s}(34)^{1_s} \right]^{2_s}$. The color part notations are $c^{AA}=\left[(12)_{\bar{3}_c}(34)_{3_c} \right]_{1_c}$, $c^{SS}=\left[(12)_{6_c}(34)_{\bar{6}_c} \right]_{1_c}$.}
\begin{tabular*}{\hsize}{@{}@{\extracolsep{\fill}}lllllll@{}}
\hline  \hline 
 $J^P$ & \multicolumn{6}{c}{ spin $\otimes$ color configurations } \\
\hline
$0^+$ 
&$\chi^{AA}_0\otimes c^{AA}$
&$\chi^{SS}_0\otimes c^{AA}$
&$\chi^{AA}_0\otimes c^{SS}$ 
&$\chi^{SS}_0\otimes c^{SS}$ & & \\
$1^+$ 
&$\chi^{SS}_1\otimes c^{AA}$
&$\chi^{SA}_1\otimes c^{AA}$
&$\chi^{AS}_1\otimes c^{AA}$
&$\chi^{SS}_1\otimes c^{SS}$
&$\chi^{SA}_1\otimes c^{SS}$
&$\chi^{AS}_1\otimes c^{SS}$\\
$2^+$ 
&$\chi^{SS}_2\otimes c^{AA}$
&$\chi^{SS}_2\otimes c^{SS}$ & & & &\\

\hline  \hline      
\end{tabular*}
\end{table*}

\section{Numerical results}~\label{sec:results}

In our simulation, we employ $1\times10^4$ walkers to implement the DMC algorithm. The ensemble undergoes $2\times10^4$ steps, each with a time increment of $\Delta t=0.01\mathrm{GeV}^{-1}$, to ensure stability. The resulting energy is averaged over the last 15000 steps to mitigate fluctuations. To estimate statistical uncertainty, we consider correlations among adjacent steps. To achieve this, we divide the steps into blocks and calculate block averages. We observe that for the four-quark systems, the blocks become uncorrelated when the block size reaches 1000 steps. Consequently, we partition the last 15000 steps into 15 blocks, each consisting of 1000 steps. We then employ the Jackknife resampling method~\cite{gattringer2009quantum} to calculate the uncertainty, which is found to be less than 1 MeV. For more details about the statistical uncertainty analysis, refer to Ref.~\cite{Ma:2022vqf}.

\subsection{Pure Constituent Quark Model}\label{subsec:pureCQM}

The calculation is performed using the AL1 and AP1 potentials. We first give the results of the $(I)J^P=(0)1^+$ system, which is the candidate of the experimental $T_{cc}$ state. A complete set of six \clabel[SpinColorSpatial]{spin-color-spatial} configurations $\left\{|T_1^0\rangle,|T_2^0\rangle,...,|T_6^0\rangle\right\}$ is included. The results are shown in Table~\ref{QQqqAll}. The binding energy is zero and -1 MeV for the AL1 and AP1 potentials respectively. With the current level of uncertainty (less than 1 MeV), it is still inconclusive whether they are below threshold or not. But the results indicate that if they exist, the binding energies are likely to be very small. 

We further calculate the ground states of all doubly heavy tetraquark systems with $J^P=0^+,1^+,2^+$. The systems with bound state solutions in AL1 and AP1 models are shown in Table~\ref{QQqqAll}. One can identify the unbound system by comparing the  Tables~\ref{QQqqAll} and \ref{threshold}. There could be resonance solutions for these systems, which will not be investigated in this work.

Comparing our results with those obtained in the same potential models using the variational approach in Ref.~\cite{Semay:1994ht}, we can find that most of our masses are lower than theirs, and more bound states are obtained. In Ref.~\cite{Meng:2020knc}, Meng \etal adopted a tuned AP1 model, using Gaussian expansion method (GEM)~\cite{Hiyama:2003cu} and obtained a binding energy of $-23$ MeV. Due to the differences in parameters, only a qualitative comparison can be made. Our results are generally consistent with theirs. In Ref.~\cite{Yang:2009zzp}, Yang \etal adopted a quark model which is very similar to the AL1 and AP1 models. Their results showed that the $(I,J)=(0,1)$ $cc\bar{n}\bar{n}$ system is unbound under the diquark-antidiquark structure and bound with a binding energy of $-1.8$ MeV under the dimeson structure. If the mixing of these two structures is taken into account, a deeper bound state of $-23.7$ MeV is obtained. For the $(I,J)=(0,1)$ $bb\bar{n}\bar{n}$ system, the diquark-antidiquark and dimeson structures give the binding energies of $-120.9$ MeV and $-11.5$ MeV respectively, and their mixing gives a deeper $-160.1$ MeV binding energy. Our results are basically consistent with their mixing picture with a loosely bound $cc\bar{n}\bar{n}$ state and a deeply bound $bb\bar{n}\bar{n}$ state. This indicates that DMC is effective even without the priori assignment of the clustering behaviors.

\subsection{Chiral Constituent Quark Model}\label{subsec:chiralQuarkModel}

As before, we first give the result of the $(I)J^P=(0)1^+$ $T_{cc}$ system. The calculation is performed using the SLM potential. A complete set of six spin-color-spatial configurations $\left\{|T_1^0\rangle,|T_2^0\rangle,...,|T_6^0\rangle\right\}$ is included. The result is shown in Table~\ref{QQqqAll}. It can be seen that the SLM yields a deeply bound $T_{cc}$ with a binding energy of $-156$ MeV. This indicates that once the spin-color configurations are complete and the spatial wave function has no presumed clustering, the $\chi$CQM will give exactly a deeply bound state instead of a shallow one near the threshold. 

Using the SLM, we further calculate the ground states of all doubly heavy tetraquark systems with $J^P=0^+,1^+,2^+$. Their corresponding dimeson thresholds are listed in Table ~\ref{threshold}. The bound states solutions are shown in Table~\ref{QQqqAll}. By comparing Tables ~\ref{threshold} and~\ref{QQqqAll}, one can identify the unbound systems.

In Ref.~\cite{Ortega:2022efc}, Ortega \etal used the same SLM model and got a very shallow bound $0(1^+)$ $cc\bar{n}\bar{n}$ state as listed in Table~\ref{QQqqAll}. This is perhaps due to the Resonating Group Method (RGM) they used. In this method, the meson wave functions are calculated in advance and input inside directly, where only the S-wave dimeson structure is considered. The higher partial wave component inside the $q\bar{q}$ pair and the distortion effect of the meson inside the tetraquark systems are neglected. This significantly restricts the wave function space and shifts a deeply bound state to the vicinity of the threshold. In the DMC calculations in this work, the spatial wave function is unclustered with no assumption of either diquark-antidiquark or dimeson structure, and all walkers diffuse freely in space. Therefore we can naturally obtain a deeply bound state. Similarly, their $T_{bb}$ bound state with quantum number $I(J^P)=0(1^+)$ is also shallower than ours. It should be noted that the corresponding threshold of the $0(0^+)$ $bb\bar{n}\bar{n}$ system actually cannot be $\bar{B}\bar{B}$, which is forbidden because of the Bose symmetry. ~\footnote{ \clabel[BBforbid]{The $\bar{B}$ meson is a boson with $I(J^P)=\frac{1}{2}(0^-)$. The total wave function of two identical $\bar{B}$ mesons has to be exchange symmetric. When constructing a $I(J^P)=0(0^+)$ $\bar{B}\bar{B}$ total wave function, the spin part $0_s\otimes 0_s\rightarrow 0_s$ is symmetric. The isospin part $\frac{1}{2}_I\otimes \frac{1}{2}_I\rightarrow 0_I$ is antisymmetric. To satisfy the exchange symmetry of two boson $\bar{B}\bar{B}$, the remaining spatial wave function has to be exchange antisymmetric, i.e., the orbital angular momentum quantum number $l$ is odd. However, the parity $P$ becomes $(-1)^l=-1$. Therefore the $0(0^+)$ $\bar{B}\bar{B}$ system is forbidden.} }

In Ref.~\cite{Deng:2022cld}, Deng \etal used a similar $\chi$CQM. In the dimeson structure with only the $1_c\times 1_c$ color configuration, a bound state with a binding energy of only $-0.34$ MeV was found. Even after adding the $8_c\times8_c$ color configuration, it remains a shallow one. For comparison, we also perform a calculation using the same model and obtain a deep binding energy of $-110$ MeV. The difference arises from the dimeson S-wave constraint in their variational approach. So essentially, it is for the same reason as described earlier about RGM. The same situation happens in Ref.~\cite{Yang:2009zzp}. Two shallow bound states with binding energies of $-0.6$ and $-0.2$ MeV are obtained using two $\chi$CQMs under the dimeson structure. 

In Ref.~\cite{Deng:2022cld}, the results of the diquark-antidiquark structure are also provided. They included two spin-color-space configurations $|T_1^0\rangle$ and $|T_2^0\rangle$, and the resulting binding energy is $-60$ MeV, shallower than our $-110$ MeV. Comparing with the configurations $\left\{|T_1^0\rangle,|T_2^0\rangle,...,|T_6^0\rangle\right\}$ included in our calculation, they only considered the two channels corresponding to the spatial diquark-antidiquark S-wave wavefunction. However, the ground state does have higher partial wave components  as explained earlier, and the remaining four channels with higher partial wave components should be included. In Ref.~\cite{Yang:2009zzp}, Yang \etal also gives the results under the diquark-antidiquark structure and under the mixing of the dimeson and diquark-antidiquark structures. Deeply bound states of $-142.4$ MeV and $-202.7$ MeV are obtained respectively. The mixing one is deeper because of the larger basis function space.

There are some other works using the GEM under the similar $\chi$CQM. Their results are qualitatively consistent with ours. In Ref.~\cite{Vijande:2003ki}, Vijande \etal used the same form of the potential as SLM, but with different parameters. They considered the spatial diquark-antidiquark structure and obtained a $-129$ MeV $cc\bar{n}\bar{n}$ bound state and a $-341$ MeV $bb\bar{n}\bar{n}$ bound state, both with $(S,I)=(1,0)$. In Ref.~\cite{Tan:2020ldi}, Tan \etal included both the dimeson and diquark-antidiquark structures and obtained deeply bound $cc\bar{n}\bar{n}$ and $bb\bar{n}\bar{n}$ states.

\begin{table*}[htp]
    \centering
    \caption{ Doubly heavy tetraquark systems with bound state solutions in AL1, AP1, and SLM models. $I$ is the isospin. $E_{\text {th
 }}$,  $E$  and $\Delta E$  are the corresponding lowest dimeson threshold, system energy, and  binding energy (in MeV) respectively in our calculations. ``NB'' represents no bound solution. We also present the $\Delta E$ in literature using the same interactions.  ``...'' represents it was not calculated in the corresponding work. }
    \label{QQqqAll}
\begin{tabular*}{\hsize}{@{}@{\extracolsep{\fill}}cccccccccccccccc@{}}
\hline 
\hline 
\multirow{2}{*}{$J^{P}$} & \multirow{2}{*}{System} & \multirow{2}{*}{$I$} & \multirow{2}{*}{Thresholds} & \multicolumn{4}{c}{AL1} & \multicolumn{4}{c}{AP1} & \multicolumn{4}{c}{SLM}\tabularnewline
\cline{5-8} \cline{9-12} \cline{13-16} 
 &  &  &  & $E_{\text{th }}$ & $E$ & $\Delta E$ & ~\cite{Semay:1994ht} & $E_{\text{th }}$ & $E$ & $\Delta E$ & ~\cite{Semay:1994ht} & $E_{\text{th }}$ & $E$ & $\Delta E$ & ~\cite{Ortega:2022efc}\tabularnewline
\hline 
\hline 
\multirow{3}{*}{$0^{+}$} & $bb\bar{n}\bar{n}$ & 1 & $\bar{B}\bar{B}$ &  &  & NB & ... &  &  & NB & ... &  &  & NB & -13.1\tabularnewline
 & $bc\bar{n}\bar{n}$ & 0 & $D\bar{B}$ & 7157 & 7136 & -21 & 1 & 7195 & 7164 & -31 & -13 & 7171 & 6986 & -185 & ...\tabularnewline
 & $bc\bar{s}\bar{n}$ & $\frac{1}{2}$ & $\bar{B}_{s}D/D_{s}\bar{B}$ &  &  & NB & ... &  &  & NB & ... & 7244 & 7243 & -1  & ...\tabularnewline
\hline 
\multirow{6}{*}{$1^{+}$} & $cc\bar{n}\bar{n}$ & 0 & $DD^{*}$ & 3880 & 3880 & 0 & 11 & 3917 & 3916 & -1 & -1 & 3915 & 3759 & -156 & -0.387\tabularnewline
 & $bb\bar{n}\bar{n}$ & 0 & $\bar{B}\bar{B}^{*}$ & 10645 & 10500 & -145  & -142 & 10680 & 10510 & -170 & -167 & 10594 & 10249 & -345 & -21.9\tabularnewline
 & $bb\bar{n}\bar{n}$ & 1 & $\bar{B}\bar{B}^{*}$ &  &  & NB & ... &  &  & NB & ... &  &  & NB & -10.5\tabularnewline
 & $bb\bar{s}\bar{n}$ & $\frac{1}{2}$ & $\bar{B}_{s}^{*}\bar{B}/\bar{B}^{*}\bar{B}_{s}$ & 10711 & 10660 & -51 & -56 & 10724 & 10667 & -57 & -61 & 10667 & 10653 & -14 & ...\tabularnewline
 & $bc\bar{n}\bar{n}$ & 0 & $D^{*}\bar{B}/D\bar{B}^{*}$ & 7214 & 7200 & -14 & -5 & 7251 & 7224 & -27 & -20 & 7215 & 7012 & -203 & ...\tabularnewline
 & $bc\bar{s}\bar{n}$ & $\frac{1}{2}$ & $D_{s}^{*}\bar{B}/D_{s}\bar{B}^{*}/D^{*}\bar{B}_{s}/D\bar{B}_{s}^{*}$ &  &  & NB & ... &  &  & NB & ... & 7291 & 7287 & -4 & ...\tabularnewline
\hline 
\multirow{2}{*}{$2^{+}$} & $bb\bar{n}\bar{n}$ & 1 & $\bar{B}^{*}\bar{B}^{*}$ &  &  & NB & 24 &  &  & NB & 4 &  &  & NB & -7.1\tabularnewline 
 & $bc\bar{n}\bar{n}$ & 0 & $D^{*}\bar{B}^{*}$ & 7368 & 7367 & -1  & ... & 7402 & 7400 & -2 & ... &  &  & NB & ...\tabularnewline

\hline 
\hline
\end{tabular*}
\end{table*}

\section{Discussion and Summary}~\label{sec:sum}

We perform a diffusion Monte Carlo (DMC) calculation of the $I(J^P)=0(1^+)$ $T_{cc}$ system and other doubly heavy tetraquark systems with $J^P=0^+,1^+,2^+$. We use two kinds of potential models, the PCQM (specifically AL1 and AP1) that contain the OGE and confinement interactions, as well as the $\chi$CQM (specifically SLM) containing the OGE, confinement, and OBE potentials. 

We include a complete set of six spin-color configurations to perform the DMC evolution with no spatial clustering assumed. For the PCQM, we are not certain if there are bound $T_{cc}$ state due to the limitation of uncertainty. But if it exists, the binding energy will be small. We further calculate the ground states of all doubly heavy tetraquark systems with $J^P=0^+,1^+,2^+$ and give the binding energies of the systems with bound state solutions. Among them, the $I(J^P)=0(0^+)$ $bc\bar{n}\bar{n}$, $0(1^+)$ $bb\bar{n}\bar{n}$, $0(1^+)$ $bc\bar{n}\bar{n}$, $\frac{1}{2}(1^+)$ $bb\bar{s}\bar{n}$, and $0(2^+)$ $bc\bar{n}\bar{n}$ systems are found to have bound states. Our results are basically consistent with the mixing picture ones using the GEM~\cite{Yang:2009zzp}, indicating that the DMC method can effectively avoid the difference in results caused by clustering in GEM.

For the $\chi$CQM case, the results show that when the color-spin-isospin configurations are complete and no spatial clustering is assumed, the SLM yields a deeply bound state with a binding energy of $-156$ MeV. This result is qualitatively consistent with the works using diquark-antidiquark structure or the mixing of dimeson and diquark-antidiquark structure in GEM. The appearance of the shallow bound states in some previous works are due to the restricted basis space. While in the DMC calculations in this work, the spatial wave function is unconstrained with no assumption of diquark-antidiquark or dimeson structure. Therefore we can obtain a deeply bound state that faithfully presents the solution of the $\chi$CQM. 

Using SLM, we further calculate the ground states of all other doubly heavy tetraquark systems with $J^P=0^+,1^+,2^+$. Among them, the $I(J^P)=0(0^+)$ $bc\bar{n}\bar{n}$,  \clabel[1/2]{$\frac{1}{2}(0^+)$} $bc\bar{s}\bar{n}$, $0(1^+)$ $bb\bar{n}\bar{n}$, $0(1^+)$ $bc\bar{n}\bar{n}$,  $\frac{1}{2}(1^+)$ $bb\bar{s}\bar{n}$, and $\frac{1}{2}(1^+)$ $bc\bar{s}\bar{n}$ systems are found to have bound states.

The $0(1^+)$ $bb\bar{n}\bar{n}$, $\frac{1}{2}(1^+)$ $bb\bar{s}\bar{n}$, $0(0^+,1^+)$ $bc\bar{n}\bar{n}$ are systems that obtain bound states in all three models. For $0(1^+)$ $bb\bar{n}\bar{n}$ and $\frac{1}{2}(1^+)$ $bb\bar{s}\bar{n}$ systems, the lattice QCD simulations consistently predicted the existence of bound states~\cite{Francis:2016hui,Junnarkar:2018twb,Meinel:2022lzo,Hudspith:2023loy}. However, for the $0(0^+,1^+)$ $bc\bar{n}\bar{n}$ system, the lattice QCD simulations did not come to the same conclusion~\cite{Francis:2018jyb,Hudspith:2020tdf,Meinel:2022lzo,Padmanath:2023rdu}.

The inability to directly obtain the experimentally observed shallow bound state $T_{cc}^+(3875)$ under the $\chi$CQM may be attributed to several reasons. The first possible one is that the parameters of the models are obtained by fitting the hadron spectrum and the hadron-hadron scattering data. So extending them directly to the tetraquark systems may no longer be applicable. The original motivation of the SLM is aimed to depict the NN scattering phase~\cite{Entem:2000mq} via RGM with presuming di-baryon clustering behaviors. If one incorporates the more general correlation of the six quarks, one perhaps obtains quite different solutions (e.g. deep bound states) instead of the NN scattering states or deuteron. The debate over whether the PCQM or $\chi$CQM is more suitable for depicting NN scattering phase shifts has persisted for many years, as exemplified in~\cite{Wang:2002ha}, but without arriving at a definitive conclusion. The preceding analysis, grounded on a single experimental result involving $T_{cc}$, does not imply the superiority of the PCQM over the $\chi$CQM.  We still have substantial room for refining the parameters of a chiral quark model that can effectively describe the tetraquark states. 

Secondly, the calculations at hadron level or under the dimeson structure usually result in shallow bound states, which indicates that there may be other mechanisms that favor the dimeson structure rather than all structures being equally likely to participate in mixing. One can understand it through the Born-Oppenheimer approximation. The potential in constituent quark model should be regarded as the Born-Oppenheimer (BO) energy in a scenario with valence quarks as slow degrees of freedom, and sea quarks and gluons as the fast degrees of freedom. Under the  BO approximation, one can first fix the positions of the valence quarks and calculate the energy of the systems, so-called BO energy, which should depend on the position of valence quarks. In the second step, the BO energy is treated as the potential of valence quarks, which is the interaction appearing in the constituent quark models. For valence quarks with different color configurations, the complicated dynamics of sea quarks and gluons are quite different, which are simplified as the flux-tubes with different topological structures. For example the di-meson configuration in Fig.~\ref{fig:structure} should be $|\bm{1}\rangle\rangle=|[(q\bar{q})_{1_{c}}(q\bar{q})_{1_{c}}]_{1_{c}}\rangle|\text{FT}_{1-1}\rangle $, where $|[(q\bar{q})_{1_{c}}(q\bar{q})_{1_{c}}]_{1_{c}}\rangle$ and $|\text{FT}_{1-1}\rangle$ represent the wave functions of valence quarks (slow degrees of freedom) and flux-tube (fast degrees of freedom) respectively. The notation $|...\rangle\rangle$ is introduced to discriminate with the naive valence quark states. In this picture, the matrix elements of $|\bm{1}\rangle\rangle$, $|\bm{1'}\rangle\rangle$, $|\bm{3}\rangle\rangle$ and $|\bm{6}\rangle\rangle$ will depend on both the valence quark wave function and flux-tube wave functions, where $|\bm{1'}\rangle\rangle$ represent another dimeson state by exchanging quarks. It is possible that the $|\bm{1}\rangle\rangle$ and $|\bm{1'}\rangle\rangle$ are favored in the tetraquark states considering the dynamics of the fast degrees of freedom~\cite{Andreev:2021eyj}. In this way, it is reasonable to incorporate the dimeson configurations in tetraquark states exclusively.  One can find a similar picture in Ref.~\cite{Wang:2023ovj}.

The recent experimental observation of tetraquark states provides us with a valuable opportunity to critically examine the popular quark models on the market and further advance them, where the diffusion Monte Carlo method without the preassignment clustering shall play a pivotal role.

\begin{appendix}

\section{Proportions of configurations}~\label{app:proportion}

\clabel[proportion]{
Due to the high computational cost required for the calculation of physical quantities related to wave functions in our current DMC coupled-channel scheme, we give the proportions of configurations using GEM as in Ref.~\cite{Meng:2023jqk}, where the included discrete quantum number configurations are the same as those in this work. The proportions of configurations for all systems with bound solutions in AL1 and SLM models are shown in Table~\ref{tab:proportionAL1} and~\ref{tab:proportionSLM} respectively. Since the AP1 results are close to those of AL1, we do not show them here. From the results we can find that the deeply bound state are often dominated by the $\overline{3}_{c}\otimes3_{c}$ component.}

\begin{table*}[htp]
    \centering
    \caption{The proportion of configurations of bound states in AL1 model. $I$ is the isospin. $E$  and $\Delta E$ are the system energy and binding energy (in MeV). }
    \label{tab:proportionAL1}
\begin{tabular*}{\hsize}{@{}@{\extracolsep{\fill}}ccccccccccc@{}}
\hline 
\hline 
$J^{P}$ & $Q_{1}Q_{2}\bar{q}_{3}\bar{q}_{4}$ & $I$  & $E$ & $\Delta E$ &\multicolumn{6}{c}{ spin $\otimes$ color configurations }  \\
\hline 
$0^{+}$&&&&& $\chi_{\overline{3}_{c}\otimes3_{c}}^{0_{s}\otimes0_{s}}$ & $\chi_{\overline{3}_{c}\otimes3_{c}}^{1_{s}\otimes1_{s}}$ & $\chi_{6_{c}\otimes\bar{6}_{c}}^{0_{s}\otimes0_{s}}$ & $\chi_{6_{c}\otimes\bar{6}_{c}}^{1_{s}\otimes1_{s}}$\tabularnewline

& $bc\bar{n}\bar{n}$ & 0 & 7136 & -21 & 38.4\% & 9.6\% & 6.9\% & 45.1\%\\
 
\hline 
$1^{+}$&&&&& $\chi_{\overline{3}_{c}\otimes3_{c}}^{1_{s}\otimes0_{s}}$ & $\chi_{\overline{3}_{c}\otimes3_{c}}^{1_{s}\otimes1_{s}}$ & $\chi_{\overline{3}_{c}\otimes3_{c}}^{0_{s}\otimes1_{s}}$ & $\chi_{6_{c}\otimes\bar{6}_{c}}^{1_{s}\otimes0_{s}}$ & $\chi_{6_{c}\otimes\bar{6}_{c}}^{1_{s}\otimes1_{s}}$ & $\chi_{6_{c}\otimes\bar{6}_{c}}^{0_{s}\otimes1_{s}}$\tabularnewline

 & $cc\bar{n}\bar{n}$ & 0 & 3880 & 0 & 49.8\% & 0.3\% & 7.1\% & 15.3\% & 0.6\% & 26.9\%\tabularnewline
 & $bb\bar{n}\bar{n}$ & 0 & 10500 & -145 & 96.4\% & 0.0\% & 0.1\% & 2.4\% & 0.0\% & 1.1\%\tabularnewline
 & $bb\bar{n}\bar{s}$ & $\frac{1}{2}$ & 10660 & -51 & 90.5\% & 0.0\% & 0.6\% & 5.3\% & 0.0\% & 3.6\%\tabularnewline
 & $bc\bar{n}\bar{n}$ & 0 & 7200 & -14 & 51.5\% & 4.7\% & 2.8\% & 7.4\% & 20.5\% & 13.1\%\tabularnewline

\hline 
$2^{+}$&&&&& $\chi_{\overline{3}_{c}\otimes3_{c}}^{1_{s}\otimes1_{s}}$ & $\chi_{6_{c}\otimes\bar{6}_{c}}^{1_{s}\otimes1_{s}}$\tabularnewline
 & $bc\bar{n}\bar{n}$ & 0 & 7367 & -1 & 28.3\% & 71.7\%\tabularnewline

\hline 
\hline
\end{tabular*}
\end{table*}

\begin{table*}[htp]
    \centering
    \caption{The proportion of configurations of bound states in SLM model. $I$ is the isospin. $E$  and $\Delta E$ are the system energy and binding energy (in MeV). }
    \label{tab:proportionSLM}
\begin{tabular*}{\hsize}{@{}@{\extracolsep{\fill}}ccccccccccc@{}}
\hline 
\hline 
$J^{P}$ & $Q_{1}Q_{2}\bar{q}_{3}\bar{q}_{4}$ & $I$  & $E$ & $\Delta E$ &\multicolumn{6}{c}{ spin $\otimes$ color configurations }  \\
\hline 
$0^{+}$&&&&& $\chi_{\overline{3}_{c}\otimes3_{c}}^{0_{s}\otimes0_{s}}$ & $\chi_{\overline{3}_{c}\otimes3_{c}}^{1_{s}\otimes1_{s}}$ & $\chi_{6_{c}\otimes\bar{6}_{c}}^{0_{s}\otimes0_{s}}$ & $\chi_{6_{c}\otimes\bar{6}_{c}}^{1_{s}\otimes1_{s}}$\tabularnewline

 & $bc\bar{n}\bar{n}$ & 0 & 6986 & -185 & 92.7\% & 0.5\% & 2.9\% & 3.9\%\tabularnewline
 & $bc\bar{n}\bar{s}$ & $\frac{1}{2}$ & 7243 & -1 & 12.0\% & 19.7\% & 11.8\% & 56.5\%\tabularnewline
 
\hline 
$1^{+}$&&&&& $\chi_{\overline{3}_{c}\otimes3_{c}}^{1_{s}\otimes0_{s}}$ & $\chi_{\overline{3}_{c}\otimes3_{c}}^{1_{s}\otimes1_{s}}$ & $\chi_{\overline{3}_{c}\otimes3_{c}}^{0_{s}\otimes1_{s}}$ & $\chi_{6_{c}\otimes\bar{6}_{c}}^{1_{s}\otimes0_{s}}$ & $\chi_{6_{c}\otimes\bar{6}_{c}}^{1_{s}\otimes1_{s}}$ & $\chi_{6_{c}\otimes\bar{6}_{c}}^{0_{s}\otimes1_{s}}$\tabularnewline

 & $cc\bar{n}\bar{n}$ & 0 & 3759 & -156 & 92.9\% & 0.1\% & 0.3\% & 4.3\% & 0.2\% & 2.2\%\tabularnewline
 & $bb\bar{n}\bar{n}$ & 0 & 10249 & -345 & 98.1\% & 0.0\% & 0.0\% & 1.7\% & 0.0\% & 0.2\%\tabularnewline
 & $bb\bar{n}\bar{s}$ & $\frac{1}{2}$ & 10653 & -14 & 72.5\% & 0.1\% & 3.0\% & 10.3\% & 0.1\% & 14.0\%\tabularnewline
 & $bc\bar{n}\bar{n}$ & 0 & 7012 & -203 & 95.2\% & 0.1\% & 0.1\% & 3.0\% & 0.7\% & 0.9\%\tabularnewline
 & $bc\bar{n}\bar{s}$ & $\frac{1}{2}$ & 7287 & -4 & 11.4\% & 13.6\% & 6.5\% & 11.3\% & 38.9\% & 18.3\%\tabularnewline

\hline 
\hline
\end{tabular*}
\end{table*}

\section{Root-mean-square radii}~\label{app:rms}

\clabel[rms]{We calculate the root-mean-square (rms) radii for all systems with bound solutions using GEM as Ref.~\cite{Meng:2023jqk}. The rms radius is defined as 
\begin{equation}
\sqrt{\left\langle r_{ij}^2\right\rangle} = \sqrt{\frac{\left\langle\psi\left|r_{i j}^2\right| \psi\right\rangle}{\langle\psi|\psi\rangle}}\,,
\end{equation}
where $|\psi\rangle$ is the calculated ground state total wave function. $r_{ij}$ is the relative distance between quark $i$ and $j$. The results of AL1 and SLM models are shown in Table~\ref{tab:rmsAL1} and~\ref{tab:rmsSLM}. We denote in the last column of the tables whether the corresponding state is a compact tetraquark state or a molecular one. We can find that the deeply bound states tend to form compact tetraquark states, and the shallow bound states tend to form molecular states. For compact tetraquark states, the heavier the masses of the quarks, the closer they are to each other.}

\begin{table*}[htbp]
    \centering
    \caption{The rms radii for bound states in AL1 model. $I$ is the isospin. $E$  and $\Delta E$ are the system energy and binding energy (in MeV). $\sqrt{\langle r_{ij}^{2}\rangle}$ is the rms radius between $i,j$ (in fm). C and M represent the compact tetraquark and molecular type respectively.}
    \label{tab:rmsAL1}
\begin{tabular*}{\hsize}{@{}@{\extracolsep{\fill}}cccccccccccc@{}}
\hline 
\hline 
$J^{P}$ & $Q_{1}Q_{2}\bar{q}_{3}\bar{q}_{4}$ & $I$ & $E$ & $\Delta E$ & $\sqrt{\langle r_{12}^{2}\rangle}$ & $\sqrt{\langle r_{13}^{2}\rangle}$ & $\sqrt{\langle r_{14}^{2}\rangle}$ & $\sqrt{\langle r_{23}^{2}\rangle}$ & $\sqrt{\langle r_{24}^{2}\rangle}$ & $\sqrt{\langle r_{34}^{2}\rangle}$ & Type   \\
\hline 
$0^{+}$ & $bc\bar{n}\bar{n}$ & 0 & 7136 & -21 & 0.77 & 0.78 & 0.78 & 0.80 & 0.80 & 1.09 & C\tabularnewline
\hline 
$1^{+}$ & $cc\bar{n}\bar{n}$ & 0 & 3880 & 0 & 0.97 & 0.83 & 0.83 & 0.83 & 0.83 & 1.14 & M\tabularnewline 
 & $bb\bar{n}\bar{n}$ & 0 & 10500 & -145 & 0.24 & 0.42 & 0.42 & 0.42 & 0.42 & 0.58 & C\tabularnewline
 & $bb\bar{n}\bar{s}$ & $\frac{1}{2}$ & 10660 & -51 & 0.32 & 0.47 & 0.39 & 0.47 & 0.39 & 0.59 & C\tabularnewline
 & $bc\bar{n}\bar{n}$ & 0 & 7200 & -14 & 0.73 & 0.78 & 0.78 & 0.80 & 0.80 & 1.09 & C\tabularnewline

\hline 
$2^{+}$ & $bc\bar{n}\bar{n}$ & 0 & 7367 & -1 & 2.08 & 1.58 & 1.58 & 1.59 & 1.59 & 2.23 & M\tabularnewline

\hline 
\hline
\end{tabular*}
\end{table*}

\begin{table*}[htbp]
    \centering
    \caption{The rms radii for bound states in SLM model. $I$ is the isospin. $E$  and $\Delta E$ are the system energy and binding energy (in MeV).  $\sqrt{\langle r_{ij}^{2}\rangle}$ is the rms radius between $i,j$ (in fm). C and M represent the compact tetraquark and molecular type respectively.}
    \label{tab:rmsSLM}
\begin{tabular*}{\hsize}{@{}@{\extracolsep{\fill}}cccccccccccc@{}}
\hline 
\hline 
$J^{P}$ & $Q_{1}Q_{2}\bar{q}_{3}\bar{q}_{4}$ & $I$ & $E$ & $\Delta E$ & $\sqrt{\langle r_{12}^{2}\rangle}$ & $\sqrt{\langle r_{13}^{2}\rangle}$ & $\sqrt{\langle r_{14}^{2}\rangle}$ & $\sqrt{\langle r_{23}^{2}\rangle}$ & $\sqrt{\langle r_{24}^{2}\rangle}$ & $\sqrt{\langle r_{34}^{2}\rangle}$ & Type   \\
\hline 
$0^{+}$  & $bc\bar{n}\bar{n}$ & 0 & 6986 & -185 & 0.28 & 0.39 & 0.39 & 0.41 & 0.41 & 0.54 & C\tabularnewline
& $bc\bar{n}\bar{s}$ & $\frac{1}{2}$ & 7243 & -1 & 1.50 & 1.52 & 0.61 & 0.71 & 1.50 & 1.62 & M\tabularnewline
\hline 
$1^{+}$ 
 & $cc\bar{n}\bar{n}$ & 0 & 3759 & -156 & 0.23 & 0.28 & 0.28 & 0.28 & 0.28 & 0.36 & C\tabularnewline
 & $bb\bar{n}\bar{n}$ & 0 & 10249 & -345 & 0.18 & 0.35 & 0.35 & 0.35 & 0.35 & 0.48 & C\tabularnewline 
 & $bb\bar{n}\bar{s}$ & $\frac{1}{2}$ & 10653 & -14 & 0.48 & 0.58 & 0.50 & 0.58 & 0.50 & 0.74 & C\tabularnewline
 & $bc\bar{n}\bar{n}$ & 0 & 7012 & -203 & 0.28 & 0.40 & 0.40 & 0.42 & 0.42 & 0.55 & C\tabularnewline 
 & $bc\bar{n}\bar{s}$ & $\frac{1}{2}$ & 7287 & -4 & 1.54 & 1.55 & 0.67 & 0.75 & 1.53 & 1.67 & M\tabularnewline

\hline 
\hline
\end{tabular*}
\end{table*}

\end{appendix}

\begin{acknowledgements}

We are grateful to Cheng-Rong Deng and Zi-Yang Lin for the helpful discussions. This project was supported by the National Natural Science Foundation of China (11975033 and 12070131001). This project was also funded by the Deutsche Forschungsgemeinschaft (DFG, German Research Foundation, Project ID 196253076-TRR 110).

\end{acknowledgements}

\bibliographystyle{apsrev4-2}
\bibliography{Ref}

\end{document}